\documentclass[aip,apl,amsmath,amssymb,reprint]{revtex4-1}
\usepackage{epsfig}
\usepackage{dsfont}
\usepackage{xcolor}
\usepackage{amsmath,bm}

\begin{document}
\title{Cascaded Brillouin lasing in monolithic barium fluoride whispering gallery mode resonators}

\author{Guoping Lin}
\email{guoping.lin@femto-st.fr}
\author{Souleymane Diallo}
\author{Khaldoun Saleh}
\author{Romain Martinenghi}
\author{Jean-Charles Beugnot}
\author{Thibaut Sylvestre}
\author{Yanne K. Chembo}
\affiliation{Optics Department, FEMTO-ST Institute [CNRS UMR6174], 25030 Besan\c con, France}


\begin{abstract}
We report the observation of stimulated Brillouin scattering and lasing at 1550~nm in barium fluoride (BaF$_2$) crystal. Brillouin lasing was achieved with ultra-high quality ($Q$) factor monolithic whispering gallery mode (WGM) mm-size disk resonators. Overmoded resonators were specifically used to provide cavity resonances for both the pump and all Brillouin Stokes waves. Single and multiple Brillouin Stokes radiations with frequency shift ranging from $8.2$~GHz up to $49$~GHz have been generated through cascaded Brillouin lasing. BaF$_2$ resonator-based Brillouin lasing can find potential applications for high-coherence lasers and microwave photonics.
\end{abstract}
\maketitle


Stimulated Brillouin scattering (SBS) is a nonlinear optical process resulting from the coherent interaction of light and acoustic waves. It is usually related to the effect of electrostriction and gives rise to inelastic light backscattering with a Doppler downshift related to the acoustic phonon frequency. Over the past years, SBS has been extensively studied in numerous optical waveguides such as optical fibers~\cite{Ippen1972Stimulated, Abedin2005Observation, Kobyakov2010Stimulated}, photonics crystal fibers~\cite{Dainese2006Stimulated, Beugnot2007Complete}, and on-chip photonic integrated circuits~\cite{Pant2011chip, Eggleton:13,Buttner2014Phase}. A variety of nonlinear materials including silica, chalcogenide or silicon have been investigated. Enhanced SBS has been recently predicted and demonstrated in nanoscale silicon photonic waveguides~\cite{Rakich2012Giant,Shin2013}, where the radiation pressure combines with electrostriction to greatly improve the Brillouin gain, thus bridging the gap between SBS and optomechanics. Enhanced and cascaded SBS can also be easily achieved in nonlinear optical cavities, leading to narrow-linewidth and efficient SBS lasing~\cite{Smith1991Narrow,Ou2014Ultra}. 

Among optical resonators, whispering gallery mode (WGM) resonators have a number of qualities that make them very attractive for investigating SBS. Their advantages include a strong light confinement due to small mode volumes and ultra-high $Q$ factors~\cite{Vahala2003Optical}. Moreover, crystalline WGM resonators are very interesting because of their broad transparency window ranging from the ultraviolet to the mid-infrared region~\cite{Lin2012High, Lin2013Wide,Wang2013Mid}. These photonic platforms thus appear as alternative and promising solutions for nonlinear applications, and thus opens an approach to harness and enhance the interaction between photons and acoustic phonons. For instance, Brillouin lasing with microwatt threshold power has recently been observed in ultra-high $Q$-factor CaF$_2$ WGM resonators~\cite{Grudinin2009Brillouin}. A narrow-linewidth Brillouin microcavity laser and an ultra-low-phase-noise microwave synthetizer have also been demonstrated using chemically etched ultrahigh-$Q$ silica-on-silicon wedge resonators~\cite{Lee2012,Li2013Microwave}. Brillouin scattering from surface acoustic waves has also been reported in MgF$_2$ WGM resonators and in silica microspheres~\cite{Savchenkov2011Surface,Tomes2009Photonic, Bahl2011Stimulated}.


\begin{figure}[b]
	\centering
	\includegraphics[width=8cm]{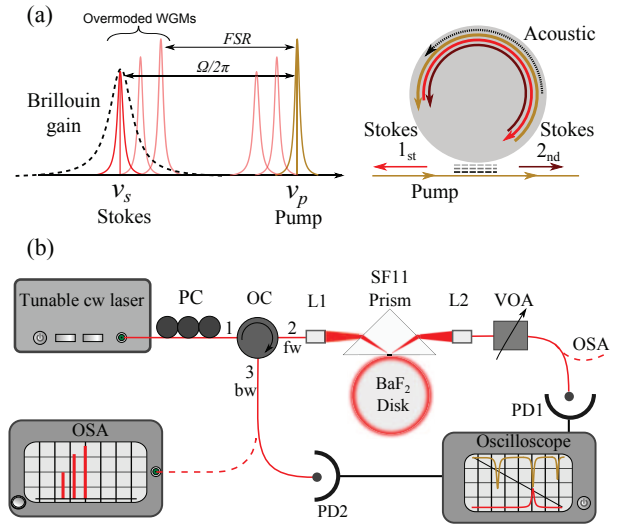}
	\caption{{\bf (a)} Left: Scheme of a Brillouin WGM resonator in a doubly resonant configuration. The dotted line designates Brillouin gain band. The solid Lorentzian lines and vertical lines show cavity resonances and laser lines, respectively; Right: Principle of a Brillouin laser based on a WGM  resonator with the incoming pump (gold), the first backward (red) and second forward (dark red) Stokes lines. {\bf (b)} Scheme of the experimental setup. PC: fiber polarization controller; OC: optical fiber circulator; L1, L2: GRIN lenses; VOA: variable optical attenuator; OSA: optical spectrum analyzer; PD1, PD2: InGaAs photodetectors; Fw: forward; Bw: backward.}
	\label{fig:schematic}
\end{figure}

In this Letter, we experimentally demonstrate Brillouin lasing at 1550 nm in ultra-high $Q$ factor WGM millimeter disk resonators based on barium fluoride (BaF$_2$) crystal~\cite{LinBarium}. We show in particular that Brillouin lasing occurs although the Brillouin frequency shift does not match the fundamental free-spectral range (FSR) of the cavity. The doubly resonant condition for the pump and the Brillouin Stokes waves is in fact fulfilled by using a group of different transverse WGMs.  Milliwatt threshold power and narrow-linewidth single and multiple Brillouin laser lines through cascaded Brillouin scattering are reported. 

The ultra-high $Q$ WGM BaF$_2$ resonators have been fabricated from commercially available disks with diameters of $12$ mm and thicknesses of $1$ mm using the mechanical polishing method.~\cite{PhysRevLett.92.043903} The polished disks then have slightly smaller diameters than the original ones. With such dimensions, it is obvious that the radiation pressure induced by light on the boundaries can be neglected. To achieve lasing effects using SBS gain, doubly resonant configurations are usually exploited, as shown schematically in Fig.~\ref{fig:schematic}(a). It is therefore needed that cavity resonances exist in both pump and Brillouin gain regimes. However, this usually requires an ultimate control on resonator geometries in the single WGM regime~\cite{Grudinin2009Brillouin,Lee2012}. The Brillouin WGM resonators studied here overcome this strict condition by using a group of transverse WGMs with different frequency offset from the fundamental FSR, as we will see thereafter. The right scheme in Fig.~\ref{fig:schematic}(a) depicts typical cascaded Brillouin scattering in a WGM disk resonator whereby a series of Stokes waves of decreasing frequencies are generated through the excitation of a circling bulk longitudinal acoustic wave. The associated density variation leads to small periodic changes of the effective refractive index along the optical resonator. When passing through this moving refractive index grating, light undergoes Bragg scattering in the backward direction according to the phase-matching condition, as in fibre-based Brillouin scattering~\cite{Ippen1972Stimulated}. The backscattered Brillouin signal undergoes a slight shift of its carrier frequency that can be expressed as~\cite{Boyd2003Nonlinear}: 
$\nu_\mathrm{B}={\Omega_B}/{2\pi}={2n_{\mathrm{eff}}V_a}/{\lambda_p}$, with $\mathrm{n_{eff}}$ being the effective refractive index of the WG optical mode, $\lambda_p$ the optical wavelength in vacuum, and $V_a$ the acoustic phase velocity. The latter can be readily calculated from the longitudinal sound speed in BaF$_2$ as $V_a=[(C_{11}+C_{12}+2C_{44})/(2\rho)]^{1/2}$, where $C_{11}$, $C_{12}$, $C_{44}$ are three independent elastic constants and $\rho$ is the material density. These values in BaF$_2$ are $ \{ 0.9199, \, 0.4157, \, 0.2568 \} \times 10^{11}$~N.m$^{-2}$ and $4.83$~g.cm$^{-3}$, respectively~\cite{Weber2002Handbook}. This gives an acoustic velocity of $V_a=4.38$~km.s$^{-1}$. If we take into account the material refractive index of $1.466$ at $1550$ nm, we find a Brillouin frequency shift $\nu_B=8.27$ GHz. Another key parameter is the Brillouin gain bandwidth $\Delta\nu_B$, which is only a few tens of MHz~\cite{Sonehara2007Frequency}. This narrow gain bandwidth implies a strict phase-matching condition for WGM Brillouin lasers, as it requires a sharp cavity resonance to coincide within the SBS gain regime, as shown in Fig.~\ref{fig:schematic}(a). As a result, Brillouin lasers were initially studied in fiber ring resonators which can feature the FSR smaller than the SBS gain bandwidth to fulfill this condition. Nevertheless, it is known that WGM disk resonators exhibit rich mode structures within one FSR (overmoded resonators)~\cite{Tomes2009Photonic, Bahl2011Stimulated}. 
 
\begin{figure}[b]
	\centering
	\includegraphics[width=7.8cm]{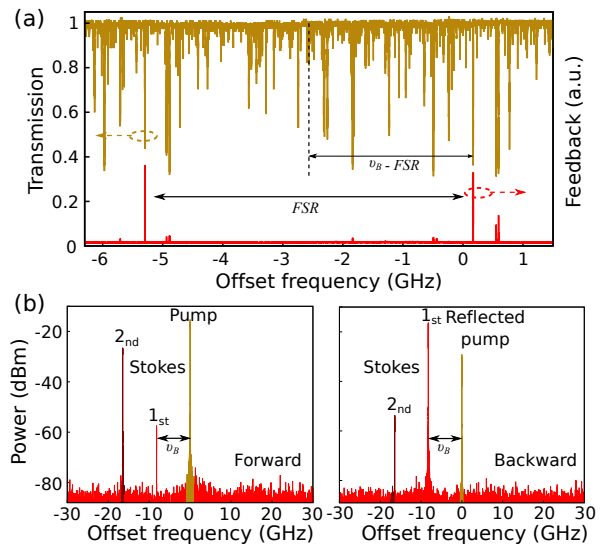}
	
	\caption{Experimental results: 
	{\bf (a)} Typical transmission spectrum of an overmoded WGM BaF$_2$ resonator with feedback Stokes Brillouin signal (red). FSR is $5.47$~GHz and $\nu_B-{\rm FSR}=2.80$~GHz.
	 {\bf (b)} Typical Brillouin spectra recorded using a high resolution optical spectrum analyzer in the forward (left) and backward (right) directions. The measured Brillouin frequency $\nu_B$ is $8.2$ GHz. Resolution is $5$~MHz.}
	\label{fig:overmoded}
\end{figure}

Figure~\ref{fig:schematic}(b) shows the experimental setup used to observe cascaded Brillouin lasing. As a pump laser, we used a tunable continuous-wave (cw) fiber laser at $1550$ nm with sub-kHz intrinsic linewidth. The laser frequency can be finely tuned by controlling the length of the fiber cavity through a piezoelectric transducer. A SF11 prism was then used to couple light into and out from the WGM BaF$_2$ resonator. A gradient-index (GRIN) lens L1 focuses the free space laser beam from the fiber port into the prism for evanescent wave coupling to WGMs. The other lens (L2) collects the reflected light signal back into a single-mode fiber for detection by a photodiode (PD1). The coupling gap between the prism and the resonator can be precisely controlled through a piezo-actuator. A fiber optical circulator was also inserted before the prism for extracting the feedback signal from the Brillouin Stokes signal into the second photodiode (PD2). The output spectra of Brillouin Stokes lines in both forward and backward direction were recorded and analyzed using a high-resolution optical spectrum analyzer (OSA, APEX 2440B, resolution down to 5 MHz).

Figure~\ref{fig:overmoded}(a) shows a typical prism-resonator transmission spectrum measured by tuning the laser frequency over more than one FSR. As it can be seen, the spectrum shows a typical multi-peak resonance structure of an overmoded WGM resonator, meaning that many higher-order transverse WGMs are resonating in the cavity. Unlike fiber ring cavities, WGMs are usually characterized with multimode behaviors because of the rich transverse mode properties in both radial and polar directions~\cite{Lin2010Excitation}. In an ultra-high $Q$ resonator, these high order transverse modes can possess similar $Q$ factors as the fundamental ones and be observed within one FSR. From Fig.~\ref{fig:overmoded}(a), we measured an FSR of 
$5.47$~GHz, in very good agreement with the theoretical value of $5.49$~GHz derived from a diameter of $d=11.87$~mm using the standard formula ${\rm FSR} = c/(\pi n_{\rm eff} d)$, where $c$ is the speed of light in vacuum. Also shown in Fig.~\ref{fig:overmoded}(a) is the feedback signal from the optical circulator simultaneously detected in PD2. Clearly, this strong feedback signal is the signature of Brillouin backscattering from the WGM resonator that match with a cavity resonance. 

To get better insight, Fig.~\ref{fig:overmoded}(b) shows both the forward and backward SBS spectra. The spectra were recorded by using the high-resolution OSA with the pump laser frequency scanning across the optical resonance. In addition to the pump frequency, we can see in both spectra the clear emergence of two Brillouin optical sidebands on the Stokes side. The frequency spacing between them is $\nu_B=8.2$~GHz, in quite good agreement with our previous estimated SBS frequency of $8.27$ GHz. The smaller experimental value could be due to the fact that the effective refractive index of a WGM is slightly smaller than the material itself. A further comparison of the two spectra of Fig.~\ref{fig:overmoded}(b) reveals that the first Brillouin Stokes sideband at $8.27$~GHz dominates in the backward spectrum whereas the second-order Brillouin line at $16.5$~GHz is mainly propagating in the forward direction. The low residual Stokes lines that appear in the opposite directions are due to weak Rayleigh scattering in the resonator~\cite{Weiss1995Splitting, Gorodetsky2000Rayleigh} and possible reflections in the setup. Figure~\ref{fig:overmoded}(b) also reveals that the first Brillouin line does not match the FSR of the resonator, whereas the second order is close to three times the FSR. 

Although the observed Brillouin frequency does not match the FSR of the resonator, we can see in Fig.~\ref{fig:overmoded}(a) the existence of WGMs with a frequency offset of $2.72$~GHz from the optical mode that has the best Stokes signal. This frequency offset matches remarkably with the Brillouin frequency shift as $\nu_B-{\rm FSR}$, indicating that higher-order transverse WGMs help to fulfill the doubly resonant condition for SBS. The threshold power is also strongly dependent on the spatial mode overlap factor $\Gamma$ between the interacting WG modes. If we assume that the optical resonance for the Brillouin Stokes waves coincide with the maximum gain position, the SBS threshold power $P_{th}$ can be expressed as~\cite{Spillane2002Ultralow}: 
\begin{equation}
  P_{\rm th}=\frac{1}{\Gamma}\frac{\pi^2n^2}{g_\mathrm{B} \lambda_p\lambda_S}\frac{V_{\mathrm{eff}}}{Q_p Q_S} 
    \label{eq:Bthreshold}
\end{equation} 
where $g_\mathrm{B}$ is the Brillouin gain of BaF$_2$ crystal, $\lambda_{P,S}$ are the wavelengths, $Q_{P,S}$ are the $Q$ factors of the pump and Stokes modes. For a resonator with its FSR close to the Brillouin frequency, the overlap factor $\Gamma$ can be close to $1$, and $\mu$W threshold behaviors can be achieved, as demonstrated in CaF$_2$~\cite{Grudinin2009Brillouin}. This condition however requires an ultimate control on resonator geometries~\cite{Grudinin2009Brillouin,Lee2012}. Here we use an overmoded WGM resonator exhibiting rich transverse modes, thus providing different frequency offsets from the fundamental FSR. As a result, we have been able to observe Brillouin lasing in three handly polished BaF$_2$ resonators with different geometries and $Q$ factors, although the threshold power is consequently higher.

\begin{figure}[t]
	\centering
	\includegraphics[width=7.5cm]{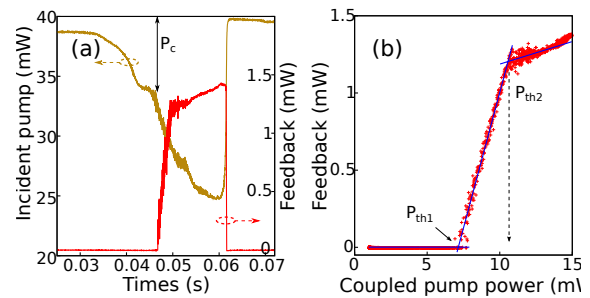}
	\caption{{\bf (a)} Experimental measurements showing the onset of stimulated Brillouin scattering in red through pump laser detuning across a thermally distorted WGM. The corresponding pump frequency scanning range is $10.5$~MHz. {\bf (b)} Brillouin Stokes power in the feedback direction as a function of the coupled pump power. $P_{\rm th1}$: $7.1$~mW; $P_{\rm th2}$: $10.6$~mW; The slope efficiency between $P_{\rm th1}$ and $P_{\rm th2}$ reaches $35\%$.}
	\label{fig:threshold}
\end{figure}
\begin{figure}[b]
	\centering
	\includegraphics[width=7.5cm]{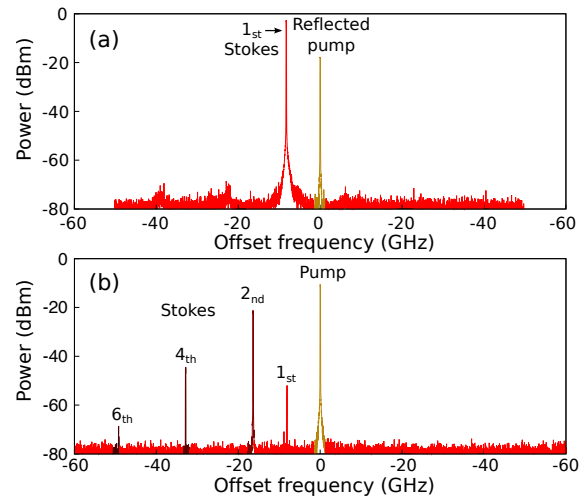}
	\caption{Experimental Brillouin spectra obtained in two different BaF$_2$ resonators: {\bf (a)}  A spectrum of a single-line WGM Brillouin laser in the backward direction. {\bf (b)} A spectrum showing cascaded Brillouin lasing up to the $6$th order recorded in the forward direction. Resolution is $100$~MHz}
	\label{fig:spectra}
\end{figure}

To further investigate the Brillouin laser performances, we carried out experiments using a fast characterization method similar to the one used for sub-microwatt threshold microlasers~\cite{Lin2012Thermal}. A WGM resonator with an intrinsic $Q$ factor of $6 \times 10^8$ was pumped with the laser frequency scanning across the resonance. The onset and the disappearance of the feedback Brillouin signal was simultaneously monitored, as shown in Fig.~\ref{fig:threshold} (a). The distorted resonance on the pump mode results from the negative thermo-optic coefficient of BaF$_2$~\cite{LinBarium}. We then plotted in Fig.~\ref{fig:threshold}(b) the Stokes signal as a function of the coupled pump power ($P_c$) to measure the laser threshold. The first-order Brillouin laser threshold was measured from a coupled pump power of $P_{\rm th1}=7.1$~mW with a slope efficiency of $35\%$. We can also see a clamping of the first Stokes laser power for a pump power of $P_{\rm th2}=10.6$~mW. This is due to the second-order Brillouin generation that circulates back into the cavity forward direction. The threshold behaviors of two Stokes lines were also confirmed in the spectrum by changing the maximum coupled pump power. We have also observed that different coupling gaps yield different threshold conditions as the loaded $Q$ factors and optical mode positions also depend on the gap.
Brillouin lasers are ideal candidates to achieve significant linewidth reduction from the pump laser source. Moreover, cascaded Brillouin lasers has been used for low-phase noise microwave generations~\cite{Li2013Microwave}. The ability to achieve single Stokes Brillouin laser and the cascaded one is therefore of uttermost importance. This function can be achieved by tailoring the geometry of the WGM resonator. The missing resonance on the second Stokes position can lead to a single Stokes laser, which has been experimentally observed in Fig.~\ref{fig:spectra} (a). On the other hand, efficient cascaded Brillouin lasing up to the $6^{\rm th}$ order is also realized as shown in Fig.~\ref{fig:spectra} (b). This was achieved because the second order Brillouin shift is very close to triple FSR ($2~\nu_B \approx 3~{\rm FSR}$) in our resonators.  
\begin{figure}[t]
	\centering
	\includegraphics[width=7.2cm]{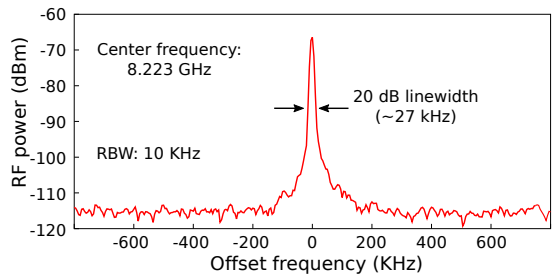}
	\caption{RF spectrum of the beat note signal, measured using a fast detector and an ESA. RBW: resolution bandwidth.}
	\label{fig:beatnote}
\end{figure}

We also examined the beat note between Brillouin Stokes lines and the pump by sending the backward signal directly into a fast photodetector. Figure~\ref{fig:beatnote} shows the corresponding RF spectrum measured using an electronic spectrum analyzer (ESA). A beat signal centered at $8.223$~GHz with a $20$~dB linewidth of $27$~kHz was observed, showing that the Brillouin linewidth of the WGMs is well below the natural linewidth in bulk fluoride (about 12 MHz)~\cite{Sonehara2007Frequency}.

In conclusion, we reported in this paper the observation of Brillouin lasing in ultra-high $Q$ BaF$_2$ WGM resonators. Both single Stokes laser and cascaded Brillouin generation have been demonstrated. The doubly resonant condition for both the pump and the Brillouin waves was fulfilled using different transverse WGMs, which release the strict condition imposed by the narrow Brillouin gain bandwidth and the cavity resonances. Our technique therefore appears as an alternative and promising solution for controlling and improving stimulated Brillouin scattering using crystalline WGM resonators and opens the way for producing efficient compact highly-coherent laser and low phase noise microwave generators for applications to telecommunications and other domains.


The authors acknowledge financial support from the European Research Council (ERC) through the projects NextPhase and Versyt.
They also acknowledge financial support from the \textit{Centre National d'Etudes Spatiales} (CNES) through the project SHYRO, from the R\'egion de Franche-Comt\'e, and from the Labex ACTION (contract No. ANR-11-LABX-0001-01).

\end{document}